\documentclass[english,preprint]{aastex}
\usepackage[T1]{fontenc}
\usepackage[latin9]{inputenc}
\usepackage{amsmath}
\usepackage{amssymb}

\makeatletter



\makeatother

\usepackage{babel}
\begin{document}

\title{Characterization of Turbulence from Submillimeter Dust Emission }

\author{Shadi Chitsazzadeh\altaffilmark{1,2}, Martin Houde \altaffilmark{2,3},
Roger H. Hildebrand \altaffilmark{4,5}, and John Vaillancourt \altaffilmark{6} }

\affil{\altaffilmark{1} Department of Physics and Astronomy, The University
of Victoria, Victoria, BC V8P 5C2, Canada}

\affil{\altaffilmark{2} Department of Physics and Astronomy, The University
of Western Ontario, London, ON N6A 3K7, Canada}

\affil{\altaffilmark{3} Division of Physics, Mathematics and Astronomy,
California Institute of Technology, Pasadena, CA 91125}

\affil{\altaffilmark{4} Department of Astronomy and Astrophysics and Enrico
Fermi Institute, The University of Chicago, Chicago, IL 60637}

\affil{\altaffilmark{5} Department of Physics, The University of Chicago,
Chicago, IL 60637}

\affil{\altaffilmark{6} Stratospheric Observatory for Infrared Astronomy,
Universities Space Research Association, NASA Ames Research Center,
Moffet Field, CA 94035, USA}
\begin{abstract}
In this paper we use our recent technique for estimating the turbulent
component of the magnetic field to derive the structure functions
of the unpolarized emission as well as that of the Stokes $Q$ and
$\mathit{U}$ parameters of the polarized emission. The solutions
for the structure functions to 350-$\mu$m SHARP polarization data
of OMC-1 allow the determination of the corresponding turbulent correlation
length scales. The estimated values for these length scales are $9\farcs4\pm0\farcs1$,
$7\farcs3\pm0\farcs1$, $12\farcs6\pm0\farcs2$ (or $20.5\pm0.2$,
$16.0\pm0.2$, and $27.5\pm0.4$ mpc at 450 pc, the adopted distance
for OMC-1) for the Stokes $\mathit{Q}$ and $\mathit{U}$ parameters,
and for the unpolarized emission $\mathit{N}$, respectively. Our
current results for $\mathit{Q}$ and $\mathit{U}$ are consistent
with previous results obtained through other methods, and may indicate
presence of anisotropy in magnetized turbulence. We infer a weak coupling
between the dust component responsible for the unpolarized emission
$\mathit{N}$ and the magnetic field $\boldsymbol{B}$ from the significant
difference between their turbulent correlation length scales.
\end{abstract}


\section{Introduction}

Magnetic fields play a crucial role in the formation of stars through
various processes, ranging from magnetic support against gravitational
collapse (see the reviews by \citealt{Shu et al. (1999)}, and \citealt{Mouschovias & Ciolek (1999)})
to magnetic braking (\citealt{Nakano (1984)}). On the other hand,
turbulence has also been suggested to be a regulating factor in the
development of the star formation process (see the reviews by \citealt{Mac Low & Klessen (2004)},
and \citealt{Elmegreen & Scalo (2004)}). The relative importance
of these two agents during the star formation process remains controversial
(e.g., \citealt{Mouschovias (2009)}; \citealt{Crutcher et al. (2010)}).

The magnetic field threading a molecular cloud is composed of a large-scale
ordered component and a turbulent (or random) one. Characterization
of the latter, e.g., through dust polarization measurements, is highly
desirable (e.g., \citealt{Lai et al. (2001)}, 2002, 2003; \citealt{Crutcher et al. (2004)}).
For example, with this information one can use the method introduced
by \citet{Chandrasekhar & Fermi (1953)} together with the estimates
of gas density and line of sight velocity dispersion in molecular
clouds (from appropriate molecular species, e.g., H$^{13}$CO$^{+}$;
see \citealt{Hildebrand et al. (2009)}, and \citealt{Houde et al. (2009)})
to derive the strength of the plane of the sky component of the magnetic
field in these regions (e.g., \citealt{Houde (2004)}; \citealt{Girart et al. (2006)};
\citealt{Curran & Chrysostomou (2007)}; \citealt{Attard et al. (2009)}).
It can also provide us with a measure of the turbulent energy content
of the gas.

A potential problem in using such a method is the necessity to make
assumptions on the structure of the large-scale component of the field.
These assumed models for the morphology of the magnetic field, even
though derived from a polarization map, may still lead to imprecise
values for the turbulent component as the assumptions regarding the
large-scale will not follow its true structure. \citet{Hildebrand et al. (2009)}
introduced a new technique where the dispersion function of the polarization
angle is used to estimate the turbulent component of the field without
making any assumptions on the morphology of the large-scale magnetic
field. This method has been studied in more details by \citet{Houde et al. (2009)},
where the effect of signal integration was taken into account in the
evaluation of the ratio of the turbulent to large-scale magnetic field
strength. This has also allowed, among other things, the determination
of the turbulent magnetic field correlation length scale. 

In this paper, we will use the methods introduced in \citet{Houde et al. (2009)}
to compare the turbulent character of the polarized emission (emanating
from the dust component that is coupled to the magnetic field) and
that of the unpolarized emission by estimating their turbulent correlation
length scales. We will start by deriving the cloud- and beam-integrated
structure functions for the Stokes $\mathit{Q}$ and $\mathit{U}$
parameters and the unpolarized emission $\mathit{N}$ in Section 2.
We will then proceed with presenting the solution for the structure
functions assuming Gaussian profiles for the beam and the turbulent
autocorrelation functions of $\mathit{Q}$, $\mathit{U}$, and $\mathit{N}$.
In Section 3, we use the previously published 350-$\mu$m SHARP polarization
map of OMC-1 of \citet{Vaillancourt et al. (2008)} to determine the
turbulent correlation lengths, and then use these results to compare
the turbulent character of the polarized emission and the unpolarized
emission and gain information on the nature of the turbulence and
magnetic field existing in the region of interest.

\section{Analysis}

\subsection{The Cloud- and Beam-integrated Structure Function of the Polarized
Components of the Emission}

The linear polarized emission is observationally determined through
the measurements of the cloud- and beam-integrated Stokes parameters

\begin{equation}
\overline{Q}(\mathbf{r})=\iint H(\mathbf{r}-\mathbf{a})\left[\frac{1}{\Delta}\intop Q\left(\mathbf{a},z\right)dz\right]d^{2}a\label{eq:1}
\end{equation}

\begin{equation}
\overline{U}(\mathbf{r})=\iint H(\mathbf{r}-\mathbf{a})\left[\frac{1}{\Delta}\intop U\left(\mathbf{a},z\right)dz\right]d^{2}a,\label{eq:2}
\end{equation}

\noindent with 

\begin{equation}
\overline{\mathit{P}}\left(\mathbf{r}\right)=\sqrt{\overline{\mathit{Q}}^{2}\left(\mathbf{r}\right)+\overline{\mathit{U}}^{2}\left(\mathbf{r}\right)}\label{eq:3}
\end{equation}

\noindent the integrated polarized emission, $H\left(\mathbf{r}\right)$
the beam profile, and $\Delta$ the maximum depth of the cloud along
any line of sight (\citealt{Houde et al. (2009)}). The unpolarized
emission $\mathit{N}$ can be estimated through the measurements of
the total emission $\mathit{I}$ and the polarized emission $\mathit{P}$
with

\noindent 
\begin{equation}
\mathit{N}\left(\mathit{\mathbf{r}}\right)=\mathit{I}\left(\mathit{\mathbf{r}}\right)-\mathit{P}\left(\mathit{\mathbf{r}}\right).\label{eq:4}
\end{equation}
In analogy to $\mathit{\overline{Q}}$ and $\mathit{\overline{U}}$,
$\overline{N}$ is defined as

\begin{equation}
\overline{N}(\mathbf{r})=\iint H(\mathbf{r}-\mathbf{a})\left[\frac{1}{\Delta}\intop N\left(\mathbf{a},z\right)dz\right]d^{2}a.\label{eq:5}
\end{equation}

\noindent The quantity for $\overline{N}\left(\mathbf{r}\right)$
defined in Equation (\ref{eq:5}) is, however, not directly measurable.
We therefore use the following approximation

\begin{equation}
\overline{N}\left(\mathbf{r}\right)\simeq\overline{I}\left(\mathbf{r}\right)-\overline{P}\left(\mathbf{r}\right).\label{eq:N_bar}
\end{equation}

\noindent The two dimensional integrals in Equations (\ref{eq:1}),
(\ref{eq:2}), and (\ref{eq:5}) are over all space. The $\mathit{z}$-axis
is along the line of sight (unit basis vector $\mathbf{e}_{z}$),
and $\mathit{\mathbf{r}}$ is the two dimensional polar radius vector
on the plane of the sky (unit basis vector $\mathbf{e}_{r}$) such
that

\begin{equation}
\mathbf{x}=r\mathbf{e}_{r}+z\mathbf{e}_{z}.\label{eq:6}
\end{equation}

\noindent Due to the mathematical similarity between the definitions
of $\mathit{\overline{Q}}$, $\mathit{\overline{U}}$, and $\overline{N}$,
the analyses will be shown for only one of them (the $\overline{Q}$
component), the results for $\mathit{\overline{U}}$ and $\mathit{\overline{N}}$
follow in a straightforward manner. 

Similarly to the characterization of the magnetic field presented
in \citet{Hildebrand et al. (2009)} and \citet{Houde et al. (2009)},
we will divide the emission - either polarized or unpolarized - into
ordered and turbulent components 

\begin{equation}
Q\left(\mathbf{a},z\right)=Q_{0}\left(\mathbf{a},z\right)+Q_{t}\left(\mathbf{a},z\right).\label{eq:7}
\end{equation}

\noindent In order to gain a quantitative estimate of the turbulent
component of the magnetic field in molecular clouds, it is necessary
to make some assumptions about the Stokes parameters ($\mathit{Q}$
and $\mathit{U}$) as well as the unpolarized emission $\mathit{N}$.
More precisely, we assume statistical independence between the ordered
and turbulent components. We will therefore have the following averages
for any two points $\mathbf{x}$ and $\mathbf{y}$

\begin{eqnarray}
\left\langle Q_{0}\left(\mathbf{x}\right)\right\rangle  & = & Q_{0}\left(\mathbf{x}\right)\nonumber \\
\left\langle Q_{t}\left(\mathbf{x}\right)\right\rangle  & = & 0\nonumber \\
\left\langle Q_{0}\left(\mathbf{x}\right)\cdot Q_{t}\left(\mathbf{y}\right)\right\rangle  & = & \left\langle Q_{0}\left(\mathbf{x}\right)\right\rangle \cdot\left\langle Q_{t}\left(\mathbf{y}\right)\right\rangle =0.\label{eq:8}
\end{eqnarray}

\noindent Moreover, we assume that $\mathit{Q}$ is stationary and
isotropic (see Equations {[}\ref{eq:14}{]} and {[}\ref{eq:15}{]}
below). Note that the distance between the two points where the Stokes
parameter $\mathit{Q}$ is measured is confined to the plane of the
sky unless otherwise stated.

The autocorrelation function of $\mathit{Q}$ can be introduced as

\begin{equation}
R_{Q}\left(v,u\right)=\left\langle Q\left(\mathbf{a},z\right)Q\left(\mathbf{a^{\prime}},z^{\prime}\right)\right\rangle \label{eq:9}
\end{equation}

\noindent with $u=\left|z^{\prime}-z\right|$ and $v=\left|\mathbf{a}^{\prime}-\mathbf{a}\right|$.
Due to the assumed statistical independence of the ordered and turbulent
components, it is possible to decompose these parts of the autocorrelation
function in the following way

\begin{equation}
R_{Q}\left(v,u\right)=R_{Q,0}\left(v,u\right)+R_{Q,t}\left(v,u\right),\label{eq:10}
\end{equation}

\noindent with

\begin{equation}
R_{Q,j}\left(v,u\right)=\left\langle Q_{j}\left(\mathbf{a},z\right)Q_{j}\left(\mathbf{a}^{\prime},z^{\prime}\right)\right\rangle \label{eq:11}
\end{equation}

\noindent in which $\mathit{j}$ stands for \textquotedbl{}0\textquotedbl{}
or \textquotedbl{}$\mathit{t}$\textquotedbl{} for the ordered and
turbulent parts, respectively.

In order to estimate the structure function of the Stokes parameter,
it is necessary to specify some characteristics of the autocorrelation
function, as well as the telescope beam profile. Following \citet{Houde et al. (2009)}
we write

\begin{equation}
R_{Q}\left(v,u\right)=R_{Q,0}\left(v,u\right)+\left\langle Q_{t}^{2}\right\rangle e^{-\left(v^{2}+u^{2}\right)/2\delta_{Q}^{2}},\label{eq:12}
\end{equation}

\noindent with $\delta_{Q}$ the correlation length scale for the
turbulent component of $\mathit{Q}$. For the time being, we assume
that $\mathit{U}$ and $N$ have turbulent correlation length scales
$\delta_{U}$ and $\delta_{N}$, which are potentially different from
$\delta_{Q}$. Moreover, these correlation lengths are both assumed
to be much smaller than the thickness of the cloud ($\Delta$). The
beam profile is also assumed Gaussian 

\begin{equation}
H\left(r\right)=\frac{1}{2\pi W^{2}}e^{-r^{2}/2W^{2}}\label{eq:13}
\end{equation}

\noindent with $\mathit{W}$ the beam radius.

For the structure function (\citealt{Falceta2008,Hildebrand et al. (2009)})
of $\mathit{\overline{Q}}$ we start with the following definition

\begin{eqnarray}
\left\langle \Delta\overline{Q}^{2}(\ell)\right\rangle  & \equiv & \left\langle \left[\overline{Q}(\mathbf{r})-\overline{Q}(\mathbf{r}+\mathbf{\mathbf{\boldsymbol{\ell}}})\right]^{2}\right\rangle \label{eq:14}\\
 & = & 2\left[\left\langle \overline{Q}^{2}\right\rangle -\left\langle \overline{Q}\,\overline{Q}(\mathbf{\ell})\right\rangle \right],\nonumber 
\end{eqnarray}

\noindent where we have used the aforementioned assumptions of isotropy
and stationarity and 

\begin{equation}
\left\langle \overline{Q}\,\overline{Q}(\mathbf{\ell})\right\rangle \equiv\left\langle \overline{Q}\left(\mathbf{r}\right)\overline{Q}(\mathbf{r}+\mathbf{\mathbf{\boldsymbol{\ell}}})\right\rangle .\label{eq:15}
\end{equation}
Under these constraints, decomposing the ordered and turbulent components
of $\mathit{Q}$, and incorporating the analytical solution provided
in Appendix A of \citet{Houde et al. (2009)} we can write

\begin{equation}
\left\langle \overline{Q}\,\overline{Q}(\mathbf{\ell})\right\rangle =\frac{\left\langle \xi\left(\ell\right)\right\rangle }{2}+\sqrt{2\pi}\left\langle Q_{t}^{2}\right\rangle \left[\frac{\delta_{Q}^{3}}{\left(\delta_{Q}^{2}+2W^{2}\right)\Delta}\right]e^{-\ell^{2}/2\left(\delta_{Q}^{2}+2W^{2}\right)},\label{eq:16}
\end{equation}

\noindent where

\begin{equation}
\left\langle \xi\left(\ell\right)\right\rangle =2\iint\iint H\left(\mathbf{a}\right)H\left(\mathbf{a}^{\prime}+\mathbf{\mathbf{\boldsymbol{\ell}}}\right)\left[\frac{2}{\Delta}\intop\left(1-\frac{u}{\Delta}\right)R_{Q,0}\left(v,u\right)du\right]d^{2}a^{\prime}d^{2}a.\label{eq:17}
\end{equation}

\noindent Inserting Equations (\ref{eq:16}) and (\ref{eq:17}) into
Equation (\ref{eq:14}) we get

\begin{eqnarray}
\left\langle \Delta\overline{Q}^{2}\left(\ell\right)\right\rangle  & = & \left[\left\langle \xi\left(0\right)\right\rangle -\left\langle \xi\left(\ell\right)\right\rangle \right]\label{eq:18}\\
 &  & +2\sqrt{2\pi}\left\langle Q_{t}^{2}\right\rangle \left[\frac{\delta_{Q}^{3}}{\left(\delta_{Q}^{2}+2W^{2}\right)\Delta}\right]\left[1-e^{-\ell^{2}/2\left(\delta_{Q}^{2}+2W^{2}\right)}\right].\nonumber 
\end{eqnarray}

\noindent The first term on the right-hand side of the above equation
(within brackets) is solely due to the ordered component of the Stokes
$\mathit{Q}$ parameter and not turbulence. This term can be expanded
as a Taylor series (similarly to the analysis presented in the Appendix
of \citealt{Houde et al. (2009)})

\begin{equation}
\left\langle \xi\left(0\right)\right\rangle -\left\langle \xi\left(\ell\right)\right\rangle =\sum_{i=1}^{\infty}c_{2i}\ell^{2i},\label{eq:19}
\end{equation}

\noindent while the isotropy in $\ell$ is incorporated by performing
the summation only on terms of even orders in $\ell$. For small displacements
satisfying $\ell\lesssim\mathit{W}$, the ordered term can be described
adequately by keeping the first order term in the Taylor series. Thus,
when $\ell\lesssim W$ we will have

\begin{eqnarray}
\left\langle \Delta\overline{Q}^{2}\left(\ell\right)\right\rangle  & \simeq & m_{Q}^{2}\ell^{2}\nonumber \\
 &  & +2\sqrt{2\pi}\left\langle Q_{t}^{2}\right\rangle \left[\frac{\delta_{Q}^{3}}{\left(\delta_{Q}^{2}+2W^{2}\right)\Delta}\right]\left[1-e^{-\ell^{2}/2\left(\delta_{Q}^{2}+2W^{2}\right)}\right],\label{eq:20}
\end{eqnarray}

\noindent where $m_{Q}^{2}=c_{2}$.

\section{Results}

We use the previously published 350-$\mu$m SHARP polarization map
of OMC-1 of \citealt{Vaillancourt et al. (2008)} to determine the
turbulent correlation lengths of the Stokes $\mathit{Q}$ and $\mathit{U}$
parameters and the unpolarized emission $\mathit{N}$. Therefore,
we refer to Equation (\ref{eq:20}) derived for the structure functions
\begin{eqnarray}
\left\langle \Delta\overline{K}^{2}\left(\ell\right)\right\rangle  & = & m_{K}^{2}\ell^{2}\nonumber \\
 &  & +2\sqrt{2\pi}\left\langle K_{t}^{2}\right\rangle \left[\frac{\delta_{K}^{3}}{\left(\delta_{K}^{2}+2W^{2}\right)\Delta}\right]\left[1-e^{-\ell^{2}/2\left(\delta_{K}^{2}+2W^{2}\right)}\right],\label{eq:21}
\end{eqnarray}

\noindent in which $\mathit{K}$ stands for $\mathit{Q}$, $\mathit{U}$,
or $\mathit{N}$. The coefficient $m_{K}^{2}$ belongs to the first
order term of the Taylor series, and $\delta_{K}$ represents the
turbulent correlation length scales of $\mathit{Q}$, $\mathit{U}$,
and $\mathit{N}$. The calculations are done using the SHARP beam
radius $W=4\farcs7$ (or FWHM = 11$\arcsec$; see \citealt{Houde et al. (2009)}). 

\begin{figure}
\plotone{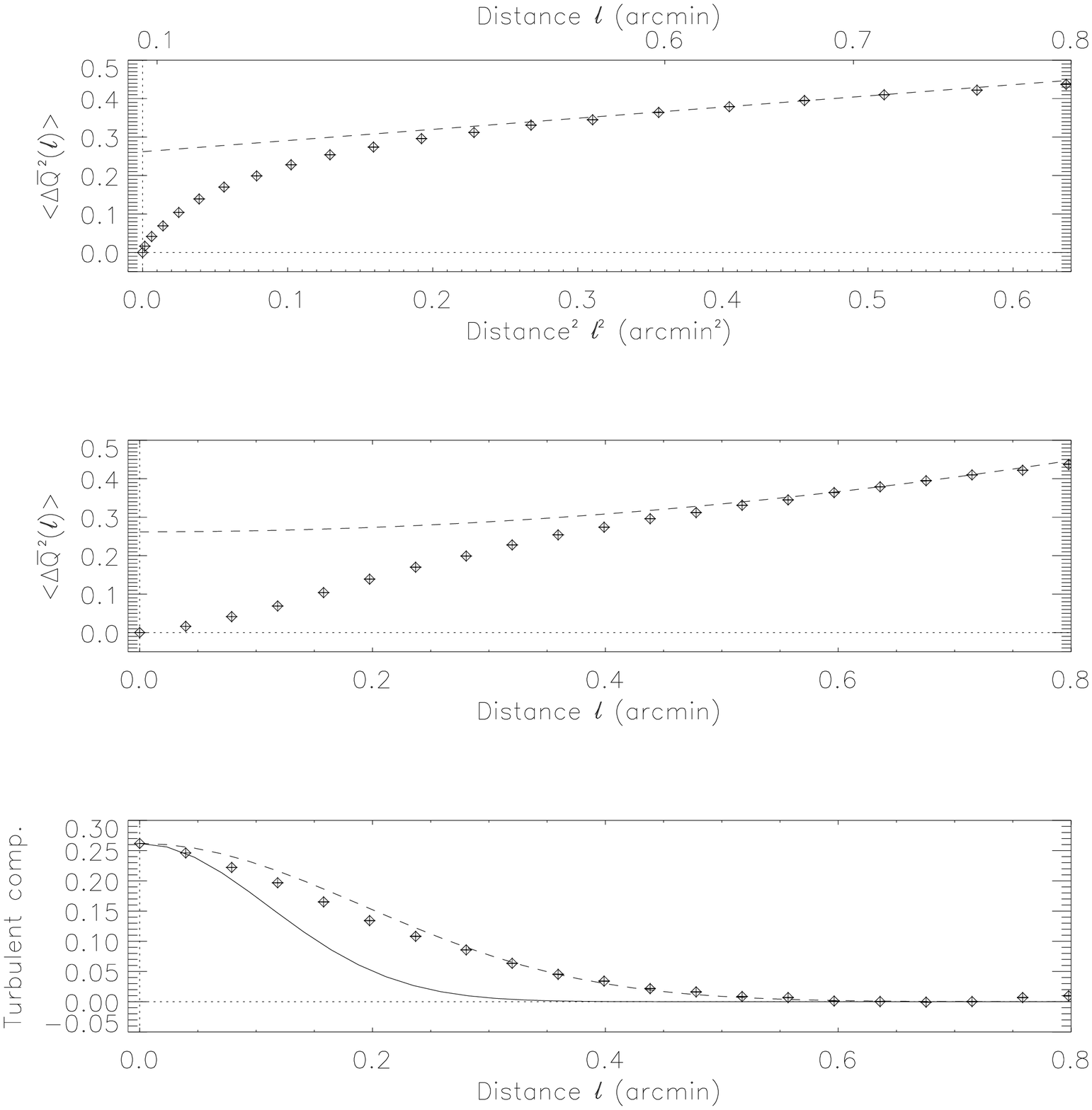}

\caption{\label{fig_Q}The structure function $\left\langle \Delta\overline{Q}^{2}\left(\ell\right)\right\rangle $
using the previously published 350-$\mu$m SHARP polarization map
of OMC-1 of \citet{Vaillancourt et al. (2008)}. \emph{Top:} Equation
(\ref{eq:23}) (broken curve) is fitted to the data (symbols) and
plotted as a function of $\ell^{2}$; \emph{middle:} similar to top
but plotted as a function of $\ell$; \emph{bottom:} the difference
between the data points and the fit of Equation (\ref{eq:23}) is
shown as symbols with the broken curve depicting the fit of Equation
(\ref{eq:22}) with $\delta_{Q}=9\farcs4\pm0\farcs1$. The solid curve
shows the contribution of the telescope beam alone (i.e., when $\delta_{K}$=
0). The telescope beam is assumed to be Gaussian.}
\end{figure}

\begin{figure}
\plotone{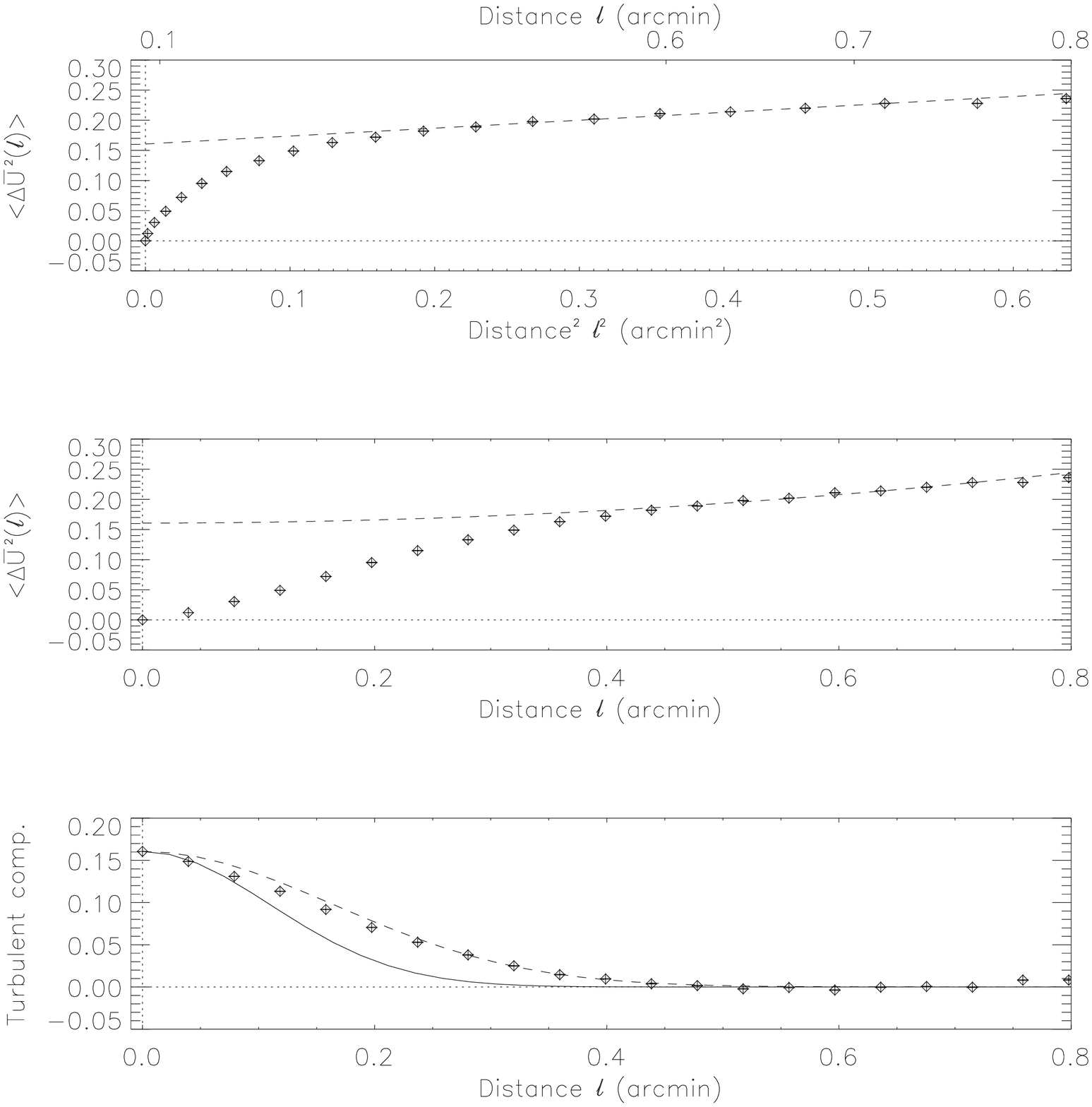}

\caption{\label{fig_U}Same as Figure 1, but for$\left\langle \Delta\overline{U}^{2}\left(\ell\right)\right\rangle $.
The turbulent correlation length is measured to be $\delta_{U}=7\farcs3\pm0\farcs1$.}
\end{figure}

\begin{figure}
\plotone{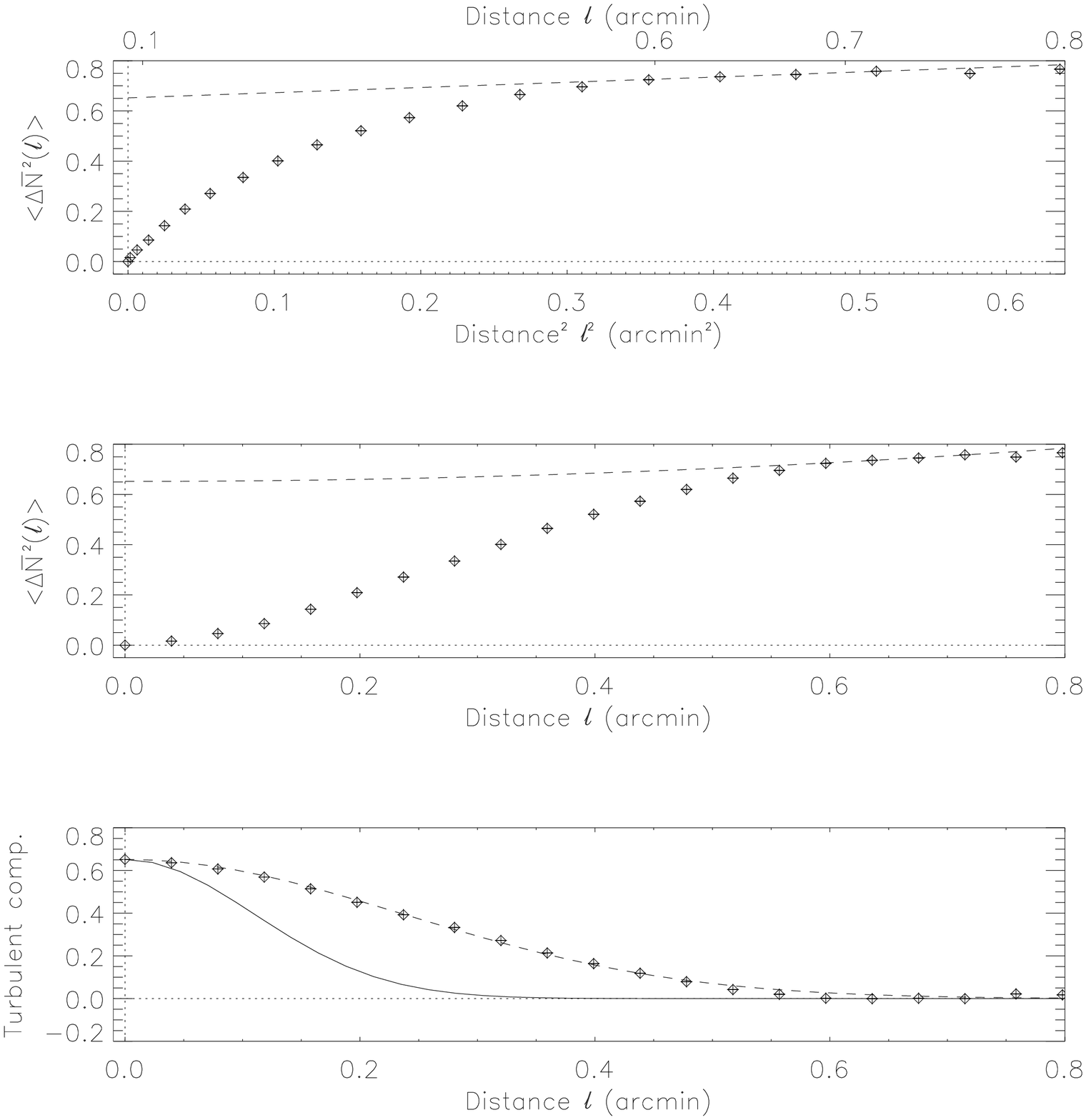}

\caption{\label{fig_N}Same as Figure 1, but for $\left\langle \Delta\overline{N}^{2}\left(\ell\right)\right\rangle $.
The turbulent correlation length is measured to be $\delta_{N}=12\farcs6\pm0\farcs2$.}
\end{figure}

Figures \ref{fig_Q}, \ref{fig_U}, and \ref{fig_N} show the results
of our fits to the data for $\mathit{Q}$, $\mathit{U}$, and $\mathit{N}$,
respectively. In all three figures, the data are shown with symbols.
In the top graphs, the broken curve does not contain the correlated
turbulent term of the function

\begin{equation}
2\sqrt{2\pi}\left\langle K_{t}^{2}\right\rangle \left[\frac{\delta_{K}^{3}}{\left(\delta_{K}^{2}+2W^{2}\right)\Delta}\right]e^{-\ell^{2}/2\left(\delta_{K}^{2}+2W^{2}\right)},\label{eq:22}
\end{equation}
but only includes the following two

\begin{equation}
m_{K}^{2}\ell^{2}+2\sqrt{2\pi}\left\langle K_{t}^{2}\right\rangle \left[\frac{\delta_{K}^{3}}{\left(\delta_{K}^{2}+2W^{2}\right)\Delta}\right].\label{eq:23}
\end{equation}
We fitted Equation (\ref{eq:23}) for values of $0\farcm5\lesssim\ell\lesssim0\farcm8$
(more details can be found in Appendix A of \citet{Houde et al. (2009)}).The
intercept of the fit at $\ell=0$ shows the level of the turbulent
component (i.e., the second term in the above equation) that can be
compared to the first term, which depicts the contribution from the
ordered part when $\ell\lesssim W$.

The middle plot shows the same information only plotted as a function
of $\ell$. We subtract the data points from the fit of Equation (\ref{eq:23})
and show the results in bottom graphs (symbols). The correlated turbulent
term (Equation {[}\ref{eq:22}{]}) is fitted to the data (broken curve)
with $\delta_{K}$ as the only fitting parameter to match the width
of the function. Even though the autocorrelation functions of $\mathit{Q}$,
$\mathit{U}$, and $\mathit{N}$ are assumed to have Gaussian patterns
and the data points in the top and middle graphs in all three figures
follow the fits quite well, it is quite unlikely that these are realistic
models for these functions. The solid curve in the bottom graphs shows
the contribution of the telescope beam alone (i.e., when $\delta_{K}$=
0 in the exponent of Equation {[}\ref{eq:22}{]}). 

The results from the fits are tabulated in Table \ref{tab:results}.
We have measured the turbulent correlation length scales of Stokes
$\mathit{Q}$ and $\mathit{U}$, and $\mathit{N}$ to be approximately
$9\farcs4\pm0\farcs1$, $7\farcs3\pm0\farcs1$, $12\farcs6\pm0\farcs2$
(or $20.5\pm0.2$, $16.0\pm0.2$, and $27.5\pm0.4$ mpc at 450 pc,
the adopted distance for OMC-1).

\begin{deluxetable}{ccc}

\tablewidth{0pt}
\tablecolumns{3}

\tablehead{
\colhead{$\delta_Q$} & \colhead{$\delta_U$} & \colhead{$\delta_N$} 
}

\tablecaption{Results from our fit of Equation (\ref{eq:23}) to the dispersion data for OMC-1.\label{tab:results}}

\startdata

$9\farcs4\pm0\farcs1$ & $7\farcs3\pm0\farcs1$ & $12\farcs6\pm0\farcs2$ \\
$20.5\pm0.2$ mpc & $16.0\pm0.2$ mpc & $27.5\pm0.4$ mpc 

\enddata

\end{deluxetable}

\section{Discussion}

The study and analysis presented in this paper seek to characterize
the magnetized turbulence in star forming regions in molecular clouds
through the determination of the turbulent correlation length scale
$\mathit{\delta}$. 

Dust grains are present in a variety of astronomical environments,
such as molecular clouds, and also tend to couple to the magnetic
field threading these regions mainly through their magnetic moment.
Comparison of the turbulent correlation length scales of different
components of the dust emission with the results from other techniques
and methods is helpful in achieving a better understanding of turbulence,
magnetic fields, and their interactions. The turbulent correlation
length scales are evaluated from the autocorrelation function of the
emission, which is the Fourier transform of the emission power spectrum,
and therefore $\delta$ being inversely proportional to the width
of the power spectrum, yields valuable information. The turbulent
correlation length scales of the Stokes parameters of the polarized
emission ($\delta_{Q}\simeq21$ mpc and $\delta_{U}\simeq16$ mpc)
are in agreement with that of the turbulent magnetic field ($\delta\simeq16$
mpc) determined by \citet{Houde et al. (2009)}.

Another comparison can be made using the results presented in \citet{Li & Houde (2008)}
for the turbulent power spectrum of coexisting ion and neutral molecular
species in M17. The spectra for these two species share the same pattern
in the inertial range, which is expected due to the tight coupling
between ions and neutrals through flux freezing. However, their spectra
cease to follow a common pattern as they decouple through turbulent
ambipolar diffusion. The values determined for $\delta_{Q}$ and $\delta_{U}$
in this paper are consistent with the analysis presented in \citet{Houde et al. (2009)},
i.e., both $\delta_{Q}$ and $\delta_{U}$ are larger than the ambipolar
diffusion scale, $\delta_{AD}$, which was recently measured to be
9.9 mpc in Orion KL by \citet{Houde et al. (2011)}. These measurements
are also consistent with more general theoretical expectations related
to turbulent ambipolar diffusion (\citealt{Lazarian et al. (2004)};
\citealt{McKee&Ostriker(2004)}; \citealt{Falceta et al. (2010)};
\citealt{Tilley et al. (2010)}) and other observational measurements
(\citet{Li & Houde (2008)}; \citealt{Hezareh et al. (2010)}).

On the other hand, there is a measurable difference between the estimated
value for $\delta_{N}$ ($\simeq28$ mpc) and the correlation length
scales of the Stokes parameters, i.e., $\delta_{Q}$ and $\delta_{U}$.
This result implies that the power spectrum of $\mathit{N}$ is different
from that of $\mathit{Q}$ and $\mathit{U}$. A lack or weakness of
coupling between the magnetic field and the dust particles responsible
for the unpolarized part of the emission can therefore be inferred
from the results.

The Stokes parameters $\mathit{Q}$ and $\mathit{U}$ are each derived
from simultaneous measurements of two orthogonal polarization states.
Considering the uncertainties in our estimates of $\delta_{Q}$ and
$\delta_{U}$, the difference between the values of these two parameters
is significant and reveals the presence of anisotropy, which is expected
for both incompressible (\citealt{Goldreich & Sridhar (1995)}; \citet{Cho et al. (2002)})
and compressible MHD turbulence (\citealt{Cho & Lazarian (2003)};
\citealt{Kowal & Lazarian (2010)}). More precisely, it is predicted
that the autocorrelation function of magnetized turbulence will have
a longer length scale in a direction parallel to the magnetic field
(as compared to an orientation perpendicular to the field). We have
measured the mean polarization angle in OMC-1 to be approximately
$30{}^{\circ}$, as can be visually verified with the polarization
map of this region presented in Figure 1(a) of \citet{Vaillancourt et al. (2008)}.
This implies that the Stokes $\mathit{U}$ parameter will be dominated
by the emission polarized (approximately) along the mean polarization
vector and therefore perpendicular to the mean magnetic field. We
thus expect that it will have a shorter correlation length scale compared
to the Stokes $\mathit{Q}$ parameter, which will more or less be
equally representative of emissions along and perpendicular to the
magnetic field. This is consistent with our results, as presented
in Table \ref{tab:results}.

Obtaining a complete turbulent power spectrum is only possible through
high resolution observations and sufficient sampling in space, which
are not available to us through polarization measurements with SHARP.
Such analyses have, however, been recently conducted by \citet{Houde et al. (2011)}.

\section{Conclusions}

In this paper we take advantage of the methods presented in \citet{Hildebrand et al. (2009)}
and \citet{Houde et al. (2009)} to determine the turbulent structure
functions of the Stokes parameters ($\mathit{Q}$ and $\mathit{U}$)
of the polarized emission and unpolarized emission $\mathit{N}$.
Subsequently, the solutions are fitted to the previously published
350-$\mu$m SHARP polarization map of OMC-1 of \citet{Vaillancourt et al. (2008)}
to estimate the turbulent correlation lengths of the Stokes parameters
$\mathit{\delta_{Q}}$ and $\mathit{\delta_{U}}$ and the unpolarized
emission $\mathit{\delta_{N}}$. Our results are consistent with the
results presented in Houde et al. (2009, 2011) and may indicate presence
of anisotropy in the magnetized turbulence. We also infer a weak coupling
between the dust component responsible for the unpolarized emission
$\mathit{N}$ and the magnetic field $\boldsymbol{B}$ from the significant
difference between their turbulent correlation length scales.

\acknowledgements{M.H.'s research is funded through the NSERC Discovery Grant, Canada
Research Chair, Canada Foundation for Innovation, Ontario Innovation
Trust, and Western's Academic Development Fund programs. The CSO is
funded through NSF AST 05-40882. This work has also been supported
in part by NSF grants AST 05-05230, AST 02-41356, and AST 05-05124.}

\clearpage

\clearpage

\clearpage

\end{document}